\documentclass[preprint]{aastex}

\def\etal{{\it et~al.~}}
\def\bsax{{\it BeppoSAX~}}
\def\ginga{{\it Ginga~}}
\def\xmm{{\it XMM-Newton~}}
\def\chandra{{\it Chandra~}}

\def\suzaku{{\it Suzaku~}}

\def\rosat{{\it ROSAT~}}

\def\rxte{{\it RXTE~}}
\def\integral{{\it Integral~}}
\def\sw{{\it Swift~}}

\def\den{~{\rm cm}^{-3}~}

\def\erg{~{\rm erg~ cm}^{-2}\ {\rm s}^{-1}~}

\begin{document}

\newcommand{\lessim}{\ \raise -2.truept\hbox{\rlap{\hbox{$\sim$}}\raise5.truept
    \hbox{$<$}\ }}

\title{Supermodel Analysis of the Hard X-Ray Excess in the Coma Cluster}

\author{R. Fusco-Femiano\altaffilmark{1}, M. Orlandini\altaffilmark{2},
M. Bonamente\altaffilmark{3,4} and A. Lapi\altaffilmark{5,6}}
\altaffiltext{1}{INAF/IASF-Roma, via del Fosso del Cavaliere,
I--00133 Roma, Italy} \altaffiltext{2}{INAF/IASF-BO, via Gobetti
101, I--40129 Bologna, Italy}\altaffiltext{3}{Department of
Physics, University of Alabama, Huntsville, AL35899}
\altaffiltext{4}{National Space Science and Technology Center,
NASA Marshall Space Flight Center, Huntsville,
AL35812}\altaffiltext{5}{Dip. Fisica, Univ. `Tor Vergata', Via
Ricerca Scientifica 1, I-00133 Roma,
Italy}\altaffiltext{6}{SISSA/ISAS, Via Bonomea 265, I-34136
Trieste, Italy}

\begin{abstract}
The Supermodel provides an accurate description of the thermal
contribution by the hot intracluster plasma which is crucial for
the analysis of the hard excess. In this paper the thermal
emissivity in the Coma cluster is derived starting from the
intracluster gas temperature and density profiles obtained by the
Supermodel analysis of X-ray observables: the \xmm temperature
profile and the \rosat brightness distribution. The Supermodel
analysis of the \bsax/PDS hard X-ray spectrum confirms our
previous results, namely an excess at the c.l. of $\sim 4.8\sigma$
and a nonthermal flux of $1.30\pm 0.40\times 10^{-11}\erg$ in the
energy range 20-80 keV. A recent joint \xmm/\suzaku analysis
reports an upper limit of $\sim 6\times 10^{-12}\erg$ in the
energy range 20-80 keV for the nonthermal flux with an average gas
temperature of 8.45$\pm$0.06 keV, and an excess of nonthermal
radiation at a confidence level above 4$\sigma$, without including
systematic effects, for an average \xmm temperature of 8.2 keV in
the \suzaku/HXD-PIN FOV, in agreement with our earlier PDS
analysis. Here we present a further evidence of the compatibility
between the \suzaku and \bsax spectra, obtained by our Supermodel
analysis of the PDS data, when the smaller size of the HXD-PIN FOV
and the two different average temperatures derived by \xmm and by
the joint \xmm/Suzaku analysis are taken into account. The
consistency of the PDS and HXD-PIN spectra reaffirms the presence
of a nonthermal component in the hard X-ray spectrum of the Coma
cluster. The Supermodel analysis of the PDS data reports an excess
at c.l. above $4\sigma$ also for the higher average temperature of
8.45 keV thanks to the PDS FOV considerably greater than the
HXD-PIN FOV.
\end{abstract}


\section{Introduction}

The Supermodel (SM) describes the density and temperature profiles
when we consider the entropy-modulated equilibrium of the
intracluster plasma (ICP) within the potential wells provided by
the dominant Dark Matter (DM). These two components are related
not only by their common potential well but also by parallel
accretion of surrounding DM and baryons into the cluster volume
(Cavaliere, Lapi \& Fusco-Femiano 2009)\footnote{The interested
reader may try for her/himself to use the fast SM algorithm made
available at the website
http://people.sissa.it/$\sim$~lapi/Supermodel/}.

An analysis of the X-ray brightness and temperature profiles for
both cool core (CC) and non cool core (NCC) clusters has been
performed in terms of the SM (Cavaliere, Lapi \& Fusco-Femiano
2009; Fusco-Femiano, Cavaliere \& Lapi 2009; thereafter FFCL09;
Lapi, Fusco-Femiano \& Cavaliere 2010). This analysis highlights
how simply the SM represents the main dichotomy "CC versus NCC"
clusters in terms of a few ICP parameters governing the radial
entropy run ($k(r) = T(r)/n(r)^{2/3}$, where $T(r)$ and $n(r)$ are
the ICP temperature and density profiles, respectively) and shows
how accurately it fits even complex brightness and temperature
profiles. This dichotomy can be represented and understood in
terms of two physical parameters marking the ICP entropy profile:
the central value $k_c$ and the outer slope $a$. More structured
temperature and brightness profiles need an additional, physical
parameter $r_f$ marking the extension of the entropy floor.

The SM has shown that the inward decline of the temperature
profile $T(r)$ characteristic of CC clusters is a feature of the
non-radiative SM equilibrium focusing also the conditions for a
cooling catastrophe that may be stabilized by ICP condensing
around and into a central massive galaxy to trigger accretion on
the nuclear black hole. The feature common to CC clusters is their
low values of $a \leq 1$ and their high values of the
concentration $c > 4$ that follow from their being old structures.
At the other extreme, the NCC clusters appear to be dynamically
young structures characterized by high values of $a$ and low
concentrations. The central flat brightness profile present in NCC
clusters like Coma and A2256 reveals large central injections of
energy and entropy deposited in the form of a floor extended out
to $r_f$. The SM challenges the complexity posed by substructures
observed in the temperature profiles of A2256 and A644. It
evidences the existence of cold regions that are remnants of a
previous cool phase partially erased by a merger event. Such cases
may be termed as RCCs for remnant of CCs. Recently, the SM
analysis (Lapi, Fusco-Femiano \& Cavaliere 2010) of the steep
temperature declines in CC clusters at low redshift (A1795 and PKS
0745-191) observed by \suzaku requires a progressive flattening of
the entropy run starting at $r \gtrsim 0.2$ of the virial radius
$R$ in agreement with the reported entropy profiles (Bautz \etal
2009; George \etal 2009). Lapi, Fusco-Femiano \& Cavaliere (2010)
argue that the entropy production at the cluster boundary is
reduced or terminated as the accretion rates of DM and
intergalactic gas peter out. This weakening of the accretion
shocks demands turbulence to develop also in the outskirts of
relaxed clusters (Cavaliere, Lapi \& Fusco-Femiano 2011).

\bsax detected the presence of non thermal (NT) radiation in
excess of the thermal ICP emission in the Coma cluster
(Fusco-Femiano \etal 1999; 2004) and Abell 2256 (Fusco-Femiano
\etal 2000; Fusco-Femiano, Landi \& Orlandini 2005). This evidence
has been claimed also by \rxte observations (Rephaeli, Gruber \&
Blanco 1999; Rephaeli \& Gruber 2002; 2003) reporting NT fluxes in
the 20-80 keV energy band in agreement with the \bsax values. The
PDS onboard \bsax was a suitable instrument to detect NT radiation
in galaxy clusters. Since clusters are very weak sources at HXR
energies above 15 keV a correct determination of the background is
crucial. Thanks to the rocking technique, the PDS was able to
perform a background measurement simultaneously with the
observations and therefore no modelling of the background was
necessary, as is required for other detectors. Moreover, the
background was very stable and low for the equatorial orbit of
\bsax.

For the Coma cluster the PDS analysis has been challenged by the
analysis of Rossetti \& Molendi (2004, thereafter RM04) with a
different software package (SAXDAS) instead of XAS. In 2007
Fusco-Femiano, Landi \& Orlandini (thereafter FFLO07) have
demonstrated that the use of the SAXDAS package allows to obtain
the same results of the previous analysis with XAS (Fusco-Femiano
\etal 2004, thereafter FF04). The main reason of the discrepancy
between the two analyses is in the non accurate selection of the
events by RM04 and not in the treatment of the total background as
reported by Wik \etal (2011). In particular, an incorrect handling
of spurious spikes due to environmental background, when present,
introduces noise that enlarges the error bars hiding the presence
of a NT excess over the thermal radiation. In fact, FFLO07 show
that the c.l. of the excess increases from 2.9$\sigma$ to
4.2$\sigma$ when the \textit{same} time windows of XAS analysis
are considered in SAXDAS. Unfortunately, this important point is
not reported in the review of Rephaeli \etal (2008). Additional
differences between the two analyses that lead to a c.l. of
4.8$\sigma$ for the excess and the remarks, including the
systematic effects, reported in RM04 are amply discussed in
FFLO07. Moreover, we underline that the consistency of the cosmic
hard X-ray background measured by \bsax/PDS (Frontera \etal 2007)
with the spectrum observed by \integral (Churazov  \etal 2007;
Bisnovatyi-Kogan \& Pozanenko 2010) and \sw (Ajello \etal 2008),
all comparable with the historic HEAO-1 measurements (Gruber \etal
1999), implies negligible PDS systematic effects as reported in
FFLO07 and Frontera \etal (2007). The correctness of the PDS
analyses is also evidenced by the agreement between the \bsax/PDS
and \suzaku/HXD-PIN spectra for the cluster Abell 3667
(Fusco-Femiano \etal 2001; Nakazawa \etal 2009).

\suzaku observations (Wik \etal 2009, thereafter W09) constrain
the thermal component by the hot ICP using a joint \xmm \&
\suzaku/HXD-PIN analysis reporting an upper limit of $\sim 6\times
10^{-12}\erg$ in the energy range 20-80 keV for the NT emission
with an average temperature of 8.45$\pm$0.06 keV. Also, they found
an excess at c.l. above 4$\sigma$ with an annular \xmm best-fit
value of 8.2 keV in the \suzaku/HXD-PIN FOV, in agreement with the
results of FF04. For the lower temperature W09 do not report the
NT flux value that this SM analysis indicates to be $\sim$ 20\%
lower than the PDS NT flux here reported for the smaller HXD-PIN
FOV than the PDS FOV (see Sect.s 2 and 3).

With our SM analysis we will show that the marginal evidence of a
NT component in the \suzaku observations is due to two combined
causes: loss of NT flux for the smaller FOV of the HXD-PIN with
respect to the \bsax/PDS and \rxte FOVs, and higher average
temperature derived by the joint analysis.

The Coma cluster has been observed also by \integral (Eckert \etal
2007; Lutovinov \etal 2008) and \sw/BAT (Ajello \etal 2009).
Eckert \etal (2007) explore the morphology of the cluster in the
HXR energy range 18-30 keV with a deep observation. The \integral
image is displaced in direction of the NGC 4839 group which is
merging with the main cluster. They associate the HXR excess in
this region with emission from a very hot region of the cluster
($T\geqslant$ 10 keV) showed by Neumann \etal (2003) in their \xmm
temperature map.

Combining data from \integral, \rxte and \rosat observatories,
Lutovinov \etal (2008) find that the thermal spectrum can be
modelled with a temperature of 8.2 keV and that the cluster is
only marginally detectable ($\sim 1.6\sigma$) in the 44-107 keV
energy band by \integral. The 20-80 keV flux of a possible NT
component ($6.0\pm 8.8\times 10^{-12}\erg$) is consistent with the
\bsax and \rxte fluxes. They also exclude with high significance
that the NT emission reported by \bsax and \rxte could be due to a
single point source.

The \sw mission is mainly devoted to detect and localize gamma-ray
bursts. \sw BAT is a coded-mask telescope optimized for the study
of point-like sources and can be used to investigate extended
objects only if these are detected as point-like. Coma instead is
\textit{extended} in BAT and part of the source flux is lost in
the BAT background. Ajello \etal (2009) treated the Coma cluster
as a point-like source considering source emission within a radius
of $\sim 10^{\prime}$ from the BAT centroid. They conclude that
the presence of a NT component arising from the cluster outskirts
cannot be excluded. More recently, Wik \etal (2011) have tested
the possibility that the difference between the \suzaku/HXD-PIN
upper limit (for $T$=8.45 keV) and the \bsax and \rxte NT fluxes
may be due to the extent of the inverse Compton (IC) emission.
Their joint \xmm/\sw BAT analysis requires an accurate
cross-calibration between the two instruments and to model both
the thermal and NT spatial distributions. Moreover, the analysis
is characterized by a higher uncertainty than for a point source
(Ajello \etal 2009; Wik \etal 2011). The derived upper limits to
the NT radiation are inconsistent with the \bsax and \rxte
observations.

In this paper, Sect. 2 reports the procedure followed for the SM
analysis of the HXR PDS spectrum in the Coma cluster. This
analysis relies on the ICP density and temperature profiles fixed
by the SM analysis (FFCL09) of the \rosat X-ray brightness and
\xmm temperature distributions. The presence of a NT spectral
component in the HXR PDS spectrum is identified by determining in
any point of the cluster the thermal ICP emissivity at a given
energy. In the previous \bsax and \rxte analyses the ICP thermal
contribution was estimated considering bremsstrahlung emission for
an average temperature in the FOV of the instruments. Sect.s 3 and
4 are devoted to the discussion and conclusions, respectively.

In our treatment, we adopt a Coma cluster redshift of 0.0232,
Hubble constant $H_0 = 70 {~\rm km~s^{-1}~Mpc^{-1}}$, and quote
errorbars at the 68\% confidence level. One arcmin corresponds to
28.12 kpc.

\section{SM Analysis of the Hard Excess}

The SM analysis of the Coma cluster (FFCL09) involves the fit to
the \xmm projected emission-weighted temperature profile (Snowden
\etal 2008) and to the \rosat surface brightness distribution
(Mohr \etal 1999) (see Fig. 1). For this paper we have performed a
slight different SM analysis of the X-ray brightness profile with
respect to that in FFCL09. We imposed the same value of $r_f =
96\pm 5$ kpc derived by the temperature profile obtaining a very
good fit to the brightness profile (see Fig. 1). Instead, a not
acceptable $\chi^2$ value is obtained imposing
$r_f=250^{+44}_{-74}$ kpc, derived by the previous analysis of the
brightness profile (see FFCL09), in the fit to the temperature
distribution. This implies that the temperature profile is more
accurate than the brightness profile in the determination of the
entropy floor extension $r_f$. From this new analysis we obtain a
value of $4.3\pm 0.4 \times 10^{-5}\den$ for the density at the
virial radius $R$. Thus, the values of $r_f$ and $n_R$ are slight
different from those reported in FFCL09, while the ICP temperature
at the virial radius is $T_R = 5.7\pm 1.0$ keV as reported in
FFCL09.

Summarizing, the free parameter values that fit the
emission-weighted temperature profile of Fig. 1 and that determine
the temperature and density profiles of Fig. 2 are: $c =
1.67^{+4.30}$ for the DM and: $k_c/k_R = (1.14\pm 0.83)\times
10^{-1}$, $a = 1.03^{+0.77}$ and $r_f/R = (4.37\pm 0.23)\times
10^{-2}$ for the ICP. These values fit the \rosat brightness
profile (see Fig. 1) when they vary within their associated
errors. The inverse process that implies to derive the parameter
values from the brightness profile to fit the temperature
distribution is not adequate for the weak dependence of the
entropy on the brightness $B$ ($k = T/n^{2/3} \sim
T^{7/6}/B^{1/3}$ where $B \sim n^2T^{1/2}$).

While the use of \rosat data is not justified in the central
regions by the PSPC angular resolution ($\sim 25^\shortparallel$)
with respect to a \xmm profile, at larger radii the latter suffers
of a greater total background. Vikhlinin \etal (2006) find an
excellent agreement between \chandra and \rosat PSPC surface
brightness data at large distances where the \rosat data allow to
have a better statistical accuracy.

Several determinations of the virial radius are given in the
literature ranging between 2 and 3 Mpc (Castander \etal 2001;
Lokas \& Mamon 2003; Kubo \etal 2007; Gavazzi \etal 2009). A value
of 2.2 Mpc (Gavazzi \etal 2009) has been adopted that corresponds
to $78.24^{\prime}$. The results of the SM analysis (FFCL09)
depend only weakly on this choice within one standard deviation.

The profiles of Fig. 1 correspond to the temperature and density
profiles of Fig. 2, with a central temperature of 9.65 keV and
central density of $2.5 \times 10^{-3}\den$. We highlight that the
projected emission-weighted temperature SM profile of Fig. 1 gives
a value of 8.21$\pm$0.08 keV (90\% c.l.) in the \suzaku FOV
($34^{\prime}\times 34^{\prime}$), the same value found by W09 in
their spectral fits to the \xmm regions of the Coma cluster and in
agreement with previous measurements. Hughes \etal (1993) derive
8.21$\pm$0.16 (90\% c.l. and including systematic effects) from
\ginga over the energy range from 1.5 to 20 keV (collimator
$1^{\circ}-2^{\circ}$ FWHM) and Arnaud \etal (2001) report
8.25$\pm$0.10 keV (90\% c.l.) in the central $r < 10^{\prime}$
region with the \xmm-EPIC-MOS detectors (energy range 0.3-10 keV).

To check the SM extrapolation to the virial radius of the
temperature profile represented by the dashed curve of Fig. 1 we
derive the average temperature within the single collimator with a
total square FOV of $65.5^{\prime}$ on a side considered by W09 to
approximate the HXD-PIN spatial response. Our SM value of $\sim$
7.85 keV is consistent with the temperature values, reported in
Table 1 of W09 in their \xmm analysis of the Coma cluster regions,
that give an average temperature of $7.79\pm 0.12$ keV. This
agreement is also visible in Fig. 3 that reports the temperature
values in the \xmm regions R1-R6 investigated by W09 showing a
temperature decline consistent with our SM extrapolation up to a
distance of more than $30^{\prime}$ ($\sim 0.4R$) from the cluster
center.

To compute the X-ray emission spectrum of the Coma cluster we
consider the MEKAL plasma model (Mewe, Gronenschild, \& van den
Oord 1985; Mewe, Lemen, \& van den Oord 1986; Kaastra 1993), the
Galactic absorption model (Morrison \& McCammon 1981), and the
abundance profile $Z(r)$ of Leccardi \etal (2010) for NCC
clusters. To take into account the temperature $T(r)$ and density
$n(r)$ profiles of Fig. 2 in SM analysis of the PDS spectrum we
utilize the routine \textit{xsmekl} outside of the \textit{XSPEC}
package. At energies above 50 keV where the MEKAL model is
undefined we use Eq. 1b of Mewe \etal (1986) to derive the photon
number emissivity at energy $E$ per unit energy interval. The
emissivity computed in any point of the cluster $F(E) =
n^2(r)\Lambda [T(r),Z(r)]$ in ${\rm photons~ cm^{-3}~s^{-1}}$ is
projected along the line-of-sight for each energy and then
integrated between a spatial interval ($r_1-r_2$) always for each
energy. $\Lambda$ is the emissivity derived by \textit{xsmekl}.
The Coma cluster flux in ${\rm photons~ cm^{-2}~s^{-1}}$
represents an \textit{additive model} that through the command
\textit{model atable} in \textit{XSPEC} fits the data.

The ICP temperature and density profiles of Fig. 2 determine the
cluster thermal emissivity in the energy range 15-80 keV. To
compare the SM spectrum with the PDS spectrum (FF04) we have
integrated the projected emissivity at a given energy in the full
\bsax FOV, $r_2 = 78^{\prime}$ ($r_1=0^{\prime}$). The SM thermal
flux at 15 keV results to be lower by a factor $\sim$1.11 than the
PDS flux at the same energy implying a NT excess even at 15 keV.
Considering the calibration offset between the \rosat/\xmm fit and
the \bsax data (see Kirsch \etal 2005) we have conservatively
normalized the SM thermal spectrum to the PDS flux observed at 15
keV. This requires a slight increase of $n_R$, at $\sim 4.5\times
10^{-5}\den$ that is within the $1\sigma$ uncertainty of the SM
determination. After the normalization, we still detect a NT
component at $E \geq$ 20 keV with significance $\sim 4.8\sigma$
with a flux in the energy range 20-80 keV of $1.30\pm 0.40\times
10^{-11}\erg$ for an assumed photon index $\Gamma$ = 2 (see Fig.
4). The significance and the flux of the NT component are
consistent with the previous analysis of FF04 ($\sigma$ = 4.8 and
flux of $1.5\pm 0.5\times 10^{-11}\erg$) using an average
temperature of 8.11 keV derived by \ginga (David \etal 1993). This
result is also in agreement with the NT component significance
greater than $4\sigma$, without including systematic effects,
obtained by \suzaku (W09) for $T$ = 8.2 keV, the same temperature
obtained by the SM for the \suzaku FOV (see Fig. 1). The NT origin
of the hard excess has been verified fitting the PDS data with a
thermal component, instead of a power law, in addition to the SM
thermal contribution. The best-fit value for the temperature is
$\sim$28.5 keV with a lower limit of $\sim$11.5 keV (90\% c.l.)
making unlikely that the hot regions reported by \xmm (Neumann
\etal 2003) and \integral (Eckert \etal 2007) observations can be
responsible for the hard tail detected by \bsax/PDS. A recent
joint analysis \xmm/\sw BAT (Wik \etal 2011) has shown
inconsistency between the NT upper limits derived for different
spatial models with the \bsax and \rxte detections. We believe
that a coded-mask telescope devoted mainly to the study of point
sources finds several difficulties to disentangle a NT component
from the ICP thermal radiation in an \textit{extended} and weak
source at HXR energies like the Coma cluster (see Ajello \etal
2009). Moreover, this analysis requires an accurate
cross-calibration between the two detectors.

\section{Discussion}

Extended radio regions observed in several galaxy clusters, radio
halos and relics, provides evidence for the presence of a
population of relativistic electrons and magnetic fields in the
ICP (see Ferrari \etal 2008). The detection of NT radiation in HXR
spectra imposes further constraints to the possible acceleration
mechanisms and origin of the relativistic electrons responsible
for NT phenomena in galaxy clusters (e.g., Brunetti \etal 2001).
The likely origin of the hard excess is IC scattering of
relativistic electrons by the cosmic microwave background (CMB)
photons. In this scenario the volume-averaged magnetic field
strength and the density of the relativistic electrons can be
determined combining radio and NT X-ray fluxes (Rephaeli 1979).
The Coma cluster exhibits a central giant radio halo and a
peripheral radio relic with total extent of about $\sim
67^{\prime}$ at 1.41 GHz with a centre $75^{\prime}$ offset with
respect to the X-ray cluster center. Besides, the very extended
($\sim 135^{\prime}$) low surface radio brightness envelope first
seen by Kronberg \etal (2007) and confirmed by Brown \& Rudnick
(2010) could be an additional source for relativistic electrons
responsible for the hard IC emission observed by \bsax and \rxte.

One of the most sensitive points in the search for NT components
is the lack of information on the thermal structure of the hot
ICP. In the analysis of the non-imaging \bsax and \rxte
observations it was only possible to consider an average
temperature in the FOV of the instruments to determine the
bremsstrahlung emissivity of the hot gas. Waiting for telescopes
able to map the HXR emission and disentangle the thermal and the
NT components (like
NuStar\footnote{http://www.nustar.caltech.edu/} and
Astro-H\footnote{http://astro-h.isas.jaxa.jp/}), we believe that
the SM is a powerful tool to constrain the ICP thermal radiation
for a confident assessment of the presence of NT spectral
components also in clusters like Coma that evidences ongoing
mergers, hallmark of a recent cluster formation. The extension of
the entropy floor, $r_f$, is interpreted in terms of the
stallation radius attained by a outbound blast triggered by a
major head-on merger or driven by a violent AGN outburst before
being degraded into adiabatic sound waves. This interpretation
relates $r_f$ to the \textit{dating} of the merger responsible for
the energy/entropy input; the good performance of the SM implies
such a time to be intermediate between the blast transit time $<$
0.1 Gyr to reach $r_f\sim 100$ kpc (see Cavaliere \& Lapi 2006)
and the time of several Gyrs needed by radiative cooling to erode
the entropy floor. Such a timing guarantees an accurate
description of the ICP thermodynamic state by the SM based on
hydrostatic equilibrium (for a more detailed discussion see Sect.
5 of FFCL09). Moreover, the equilibrium of the ICP is somewhat
faster to attain the DM's (see Ricker \& Sarazin 2001; Lapi \etal
2005) and the circularized data (integrated over annuli; see
Snowden \etal 2008) tend to smooth out local, limited deviations
from spherical hydrostatics and to better agree with equilibrium.
Conditions of disequilibrium are present in clusters like the
Bullet cluster (see Clowe \etal 2006) or MACS J0025.4-1222 (see
Brada$\check{{\rm c}}$ \etal 2008). These conditions, due to
stronger if rarer energy injections by deep major mergers, prevent
a SM description of the X-ray observables.

The SM allows to determine more correctly the thermal ICP
contribution than the temperature maps that are limited in
extension ($\lesssim R/2$), except for a handful of clusters
observed by \suzaku (Bautz \etal 2009; George \etal 2009; Reiprich
\etal 2009; Kawaharada \etal 2010; Hoshino \etal 2010) that does
not include Coma, while the \bsax FOV extends to $\sim R$.
Finally, the accurate SM fit to the brightness profile (see Fig.
1) implies that the relevance of the cluster ellipticity is mild.

To evidence the presence of a NT feature, we considered the SM ICP
temperature and density profiles to derive the underlying
contribution of the hot ICP to the HXR Coma spectrum observed by
\bsax/PDS in the energy range 15-80 keV. The profiles of Fig. 2
have been obtained by the SM analysis of the X-ray observables of
the Coma cluster: the \xmm projected emission-weighted temperature
(Snowden \etal 2008) and the \rosat brightness distribution in the
energy range 0.5-2 keV (Mohr \etal 1999). Notice that the SM
extrapolation of the temperature profile is in agreement with the
more recent analysis of the \xmm data by W09 up to a distance of
$\sim 30^{\prime}$ from the cluster center (see Fig. 3), lending
additional support to the use of the SM profiles. A further check
of the validity of the SM profiles of Fig. 2 is given by the fit
to the \rosat PSPC spectrum (energy range 1-2 keV) in the region
$20^{\prime}-40^{\prime}$ around the center of the Coma cluster
(Bonamente \etal 2003). The SM spectrum is lower than the \rosat
data by only a factor $\sim$ 1.06 (see Fig. 5).

The joint \xmm/\suzaku HXD-PIN analysis (W09) derives a mean
temperature of 8.45 keV in the HXD-PIN FOV greater than the
temperature used in the \bsax and \rxte analyses. The authors
suggest that the lower temperature may be the effect of the larger
FOVs of the two X-ray detectors that include emission from more
cool gas in the cluster outskirts. This emission that lowers the
average temperature is determined mainly by emission at energies
E$<$10 keV. But a distribution of higher than average temperature
regions can increase the average gas temperature observed at high
energies. These regions with $T \gtrsim 10$ keV  are observed also
by \xmm (Neumann \etal 2003) and \integral (Eckert \etal 2007).

With reference to this interpretation, we notice that the value of
8.45 keV in the HXD-PIN implies a temperature run that appears to
be in disagreement with the \xmm profile (Snowden \etal 2008; W09)
as shown by the dot-dashed curve in Fig. 3. Moreover, as reported
in the Introduction a temperature of 8.2 keV has been derived
combining data from \integral, \rxte and \rosat (Lutovinov \etal
2008).

An alternative and more likely explanation for the higher
temperature value found in the joint \xmm/HXD-PIN analysis may be
given by the presence of the NT component itself in the spectrum
which is responsible for the increase of the average temperature.
A power law component in fact raises the exponential decline of
the thermal emission, resulting in a higher best-fit thermal
temperature. The poor fit with $T$ = 8.2 keV relative to the fit
in which the temperature is a free parameter may be indicative of
the presence of a second component in the Coma spectrum mainly
\textit{visible }in the energy range covered by the HXD-PIN data.

This SM analysis of the HXR spectrum in the Coma cluster confirms
the results of the previous analysis by FF04. However, to remove
the possibility that the existence of a NT excess in the HXR
spectrum of the Coma cluster may be tied to the ICP average
temperature value, we have considered the temperature profile
(dot-dashed line in Fig. 3) that gives the average temperature of
8.45 keV in the HXD-PIN FOV as reported by W09 in their joint
\xmm/Suzaku analysis. In this case, the PDS spectrum reports a
c.l. for the excess of $\sim 4.3\sigma$ and a NT flux of $1.15\pm
0.41\times 10^{-11}\erg$ in the energy range 20-80 keV with
$\Gamma$=2.

We examine also the possibility that the existence of the NT
excess may depend on the temperature profile in the spatial range
between $30^{\prime}$ and the virial radius ($R=78.24^{\prime}$),
up to now not covered by observations. To be conservative we have
considered a flat temperature profile in this interval with a
constant temperature of $\sim$7.8 keV that is the value at
$r=30^{\prime}\simeq 0.4R$ (see Fig. 2). Although the flattening
of the temperature profile seems to be very unlikely (see Fig. 3),
in this case the excess is at the c.l. of $\sim 4.4\sigma$.
Finally, we have considered both these two effects: the higher
average temperature of 8.45 keV and the flat temperature profile
at $r\geq 30^{\prime}$ with a constant temperature value of
$\sim$8 keV. Also in these conditions the NT excess does not
vanish in the HXR PDS spectrum though at the level of $\sim
3.8\sigma$.

Independently from the real average temperature in the HXD-PIN FOV
a relevant point emerges from the analysis of W09. They report
that with a \xmm average temperature of 8.2 keV, a nonthermal
excess with c.l. greater than 4$\sigma$ is present in the \suzaku
data, without including systematic effects, assuming a fixed
photon index $\Gamma$ = 2 for the power law component. This
result, absolutely in agreement with the PDS analysis (FF04),
implies that the HXD-PIN spectrum is consistent with the PDS
spectrum and therefore in disagreement with the PDS spectrum of
RM04 that found a very marginal c.l. for the excess using the same
temperature (8.21 keV) and without considering systematic effects.

We also address the agreement between the \suzaku and \bsax
spectra with the following tests: \textit{a)} we consider the
smaller FOV of \suzaku HXD-PIN with respect to that of the PDS and
temperature profile for an average $T$ = 8.2 keV (continuous and
dashed lines in Fig. 1); in this case, we obtain a NT flux in the
20-80 keV energy band of $1.05\pm 0.41\times 10^{-11}\erg$ which
is $\sim 20\%$ lower than the PDS NT flux of $1.30\pm 0.40\times
10^{-11}\erg$ (see Sect. 2). The exclusion of part of the cooler
cluster regions due to the smaller HXD-PIN FOV reduces the ICP
contribution to the thermal emission mainly at energies around 15
keV. This determines a flatter ICP thermal spectrum and thus a
lower NT excess that is at a c.l. of $\sim 4\sigma$ in agreement
with the W09 analysis.

\textit{b)} we use the \suzaku FOV and temperature profile that
gives an average $T$ of 8.45 keV (dot-dashed line in Fig. 3). In
this case, the NT flux is $8.7\pm 4.2\times 10^{-12}\erg$ with a
decrease of $\sim 33\%$ with respect the PDS flux due to the
smaller HXD-PIN FOV and the higher average temperature. This flux
value is consistent with the upper limit reported by W09 of
$6\times 10^{-12}\erg$ for the NT spectral component with $\Gamma$
= 2.

This agreement between the \suzaku NT upper limit and the \bsax
detection has a further confirmation by Fig. 8 of Wik \etal (2011)
when different spatial models are examined.

We believe that the NT flux of $1.05\pm 0.41\times 10^{-11}\erg$
(see case \textit{a)}) cannot be much different from the value
measured by W09 for T=8.2 keV, and not reported in their paper
(only the c.l. of the excess is in W09), because the flux
determined for the case \textit{b)} is consistent with the \suzaku
upper limit.

Thus our SM analysis of the PDS spectrum reproduces the two
results present in the W09 analysis: an excess at the c.l. of
$\sim 4\sigma$ for an average \xmm temperature of 8.2 keV, and the
upper limit for the NT flux with an average temperature of 8.45
keV obtained by the joint \xmm/\suzaku analysis. All this
reinforces the consistency of the PDS and HXD-PIN spectra and
therefore the presence of an hard tail in the Coma cluster
spectrum.

\section{Conclusions}

For the first time the HXR spectrum of the Coma cluster has been
analyzed using ICP temperature and density profiles instead to
consider bremsstrahlung emission for an average temperature in the
detector FOVs. These profiles are determined by the SM analysis of
X-ray observables. This procedure has allowed to obtain further
checks on the relevant point present in the \suzaku analysis that
for a \xmm temperature $T$= 8.2 keV reports a NT excess at c.l.
$\gtrsim 4\sigma$ absolutely consistent with the results of FF04
(and therefore in disagreement with those of RM04). We have shown
that the compatibility between the PDS and HXD-PIN spectra has a
robust cross-check when in our SM analysis of the PDS data we
consider the smaller HXD-PIN FOV and the different average
temperatures of 8.2 keV and 8.45 keV (cases \textit{a} and
\textit{b}, respectively). The agreement between the two spectra
is a further confirmation of the presence in the Coma cluster of a
NT component. As reported in the previous Section, the PDS
spectrum gives a hard excess with significance above 4$\sigma$
\textit{also} for an ICP average temperature of 8.45 keV. This
detection by the PDS is possible thanks to its FOV, a factor
$\sim$ 4 greater than the HXD-PIN FOV that instead reports a flux
upper limit for the same temperature.

\begin{acknowledgements}
We thank Alfonso Cavaliere, Gianfranco Brunetti, Fabio
Gastaldello, Francesco Lazzarotto and Fabio Pizzolato for helpful
discussions and our referee for insightful comments, helpful
toward improving our presentation. A.L. thanks SISSA for warm
hospitality.
\end{acknowledgements}

\clearpage

\begin{figure*}
\centering \rotatebox{-00}{\epsscale{1.0}\plottwo{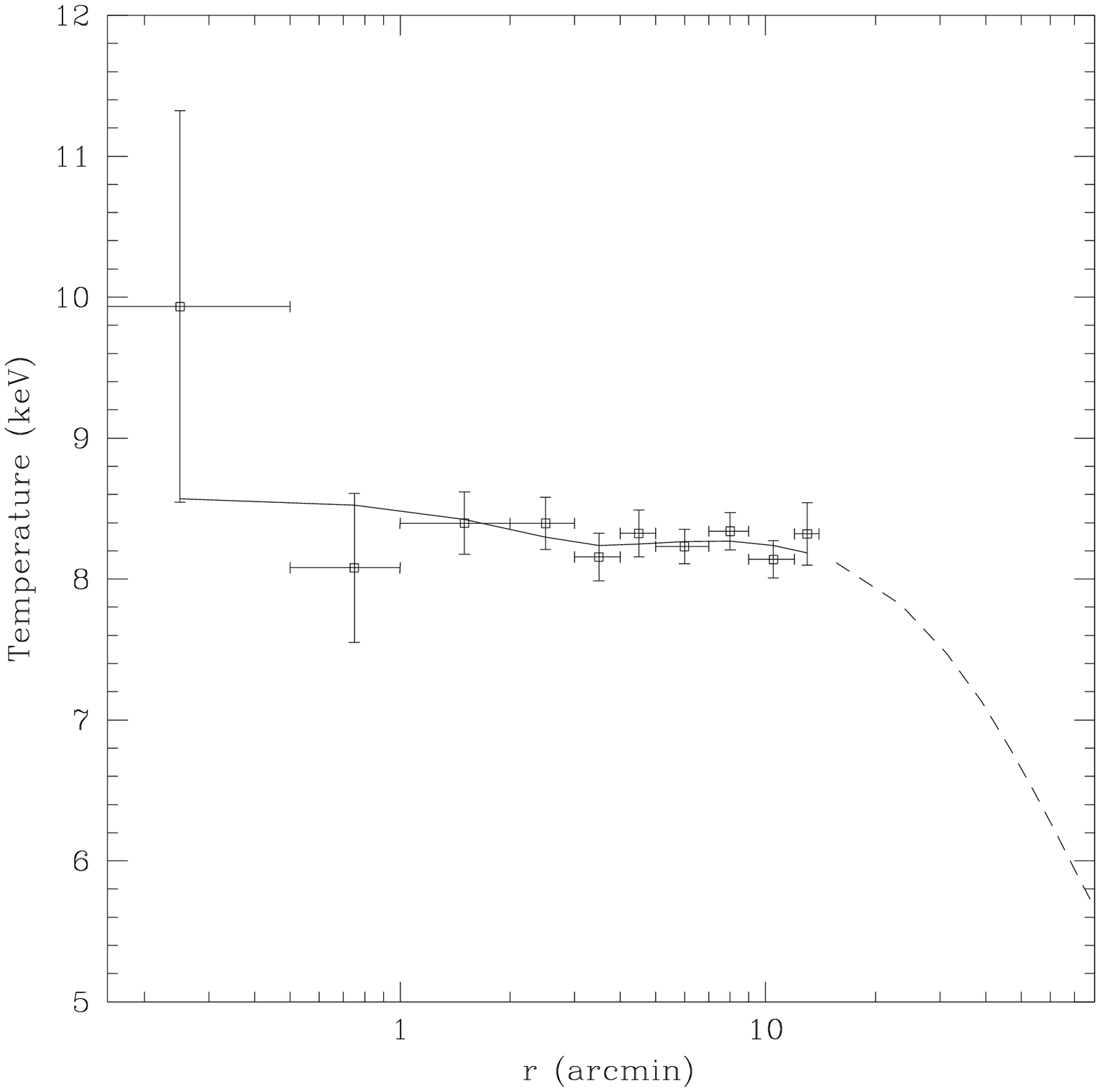}{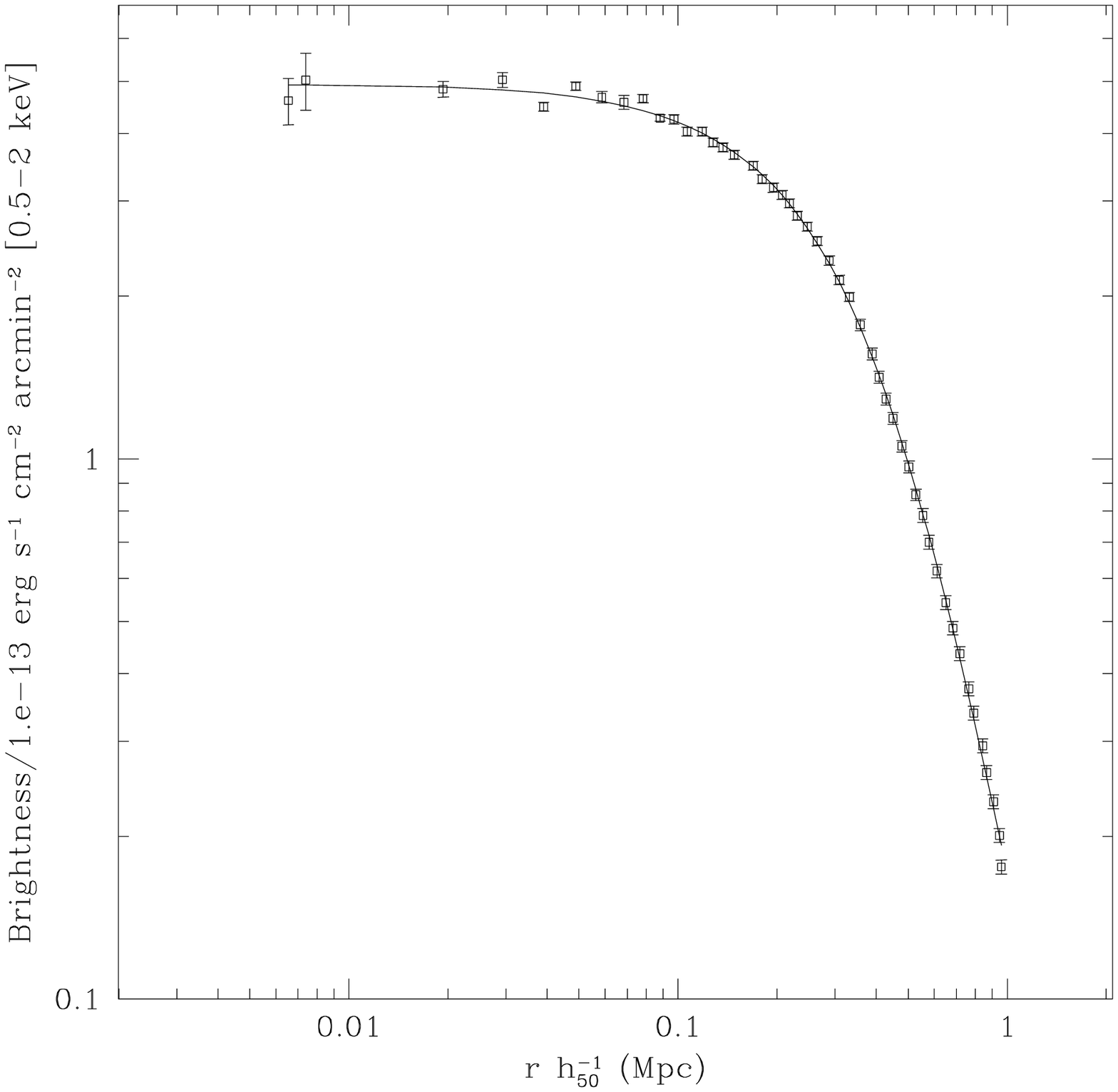}}
\caption{\textit{left panel}: Projected emission-weighted
temperature profile measured by \xmm (Snowden \etal 2008). The
continuous line represents the SM fit (see FFCL09). The dashed
line is the extrapolation of the SM fit to the virial radius $R$.
This profile gives an average temperature of 8.21$\pm$0.08 keV
(90\% c.l.) in the HXD-PIN FOV ($34^{\prime}\times 34^{\prime}$),
the same value reported by W09 in their \xmm analysis of the Coma
cluster; \textit{right panel}: Brightness profile in the energy
range 0.5-2 keV measured by \rosat (Mohr \etal 1999). The
continuous line represents the SM fit ($\chi^2$=55.5/44).
}\label{fig1}
\end{figure*}



\clearpage

\begin{figure*}
\centering \rotatebox{-00}{\epsscale{1.0}
\plottwo{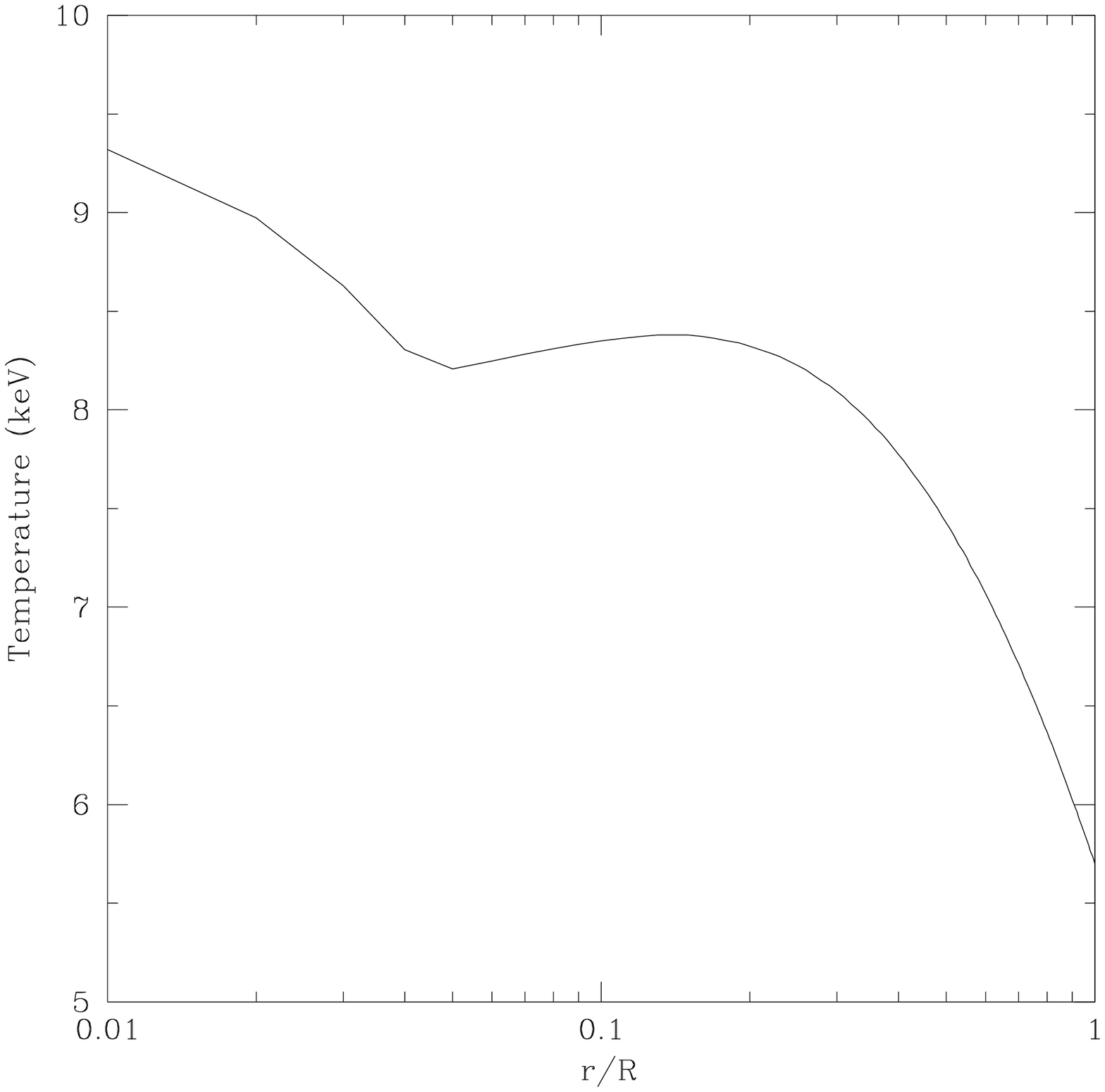}{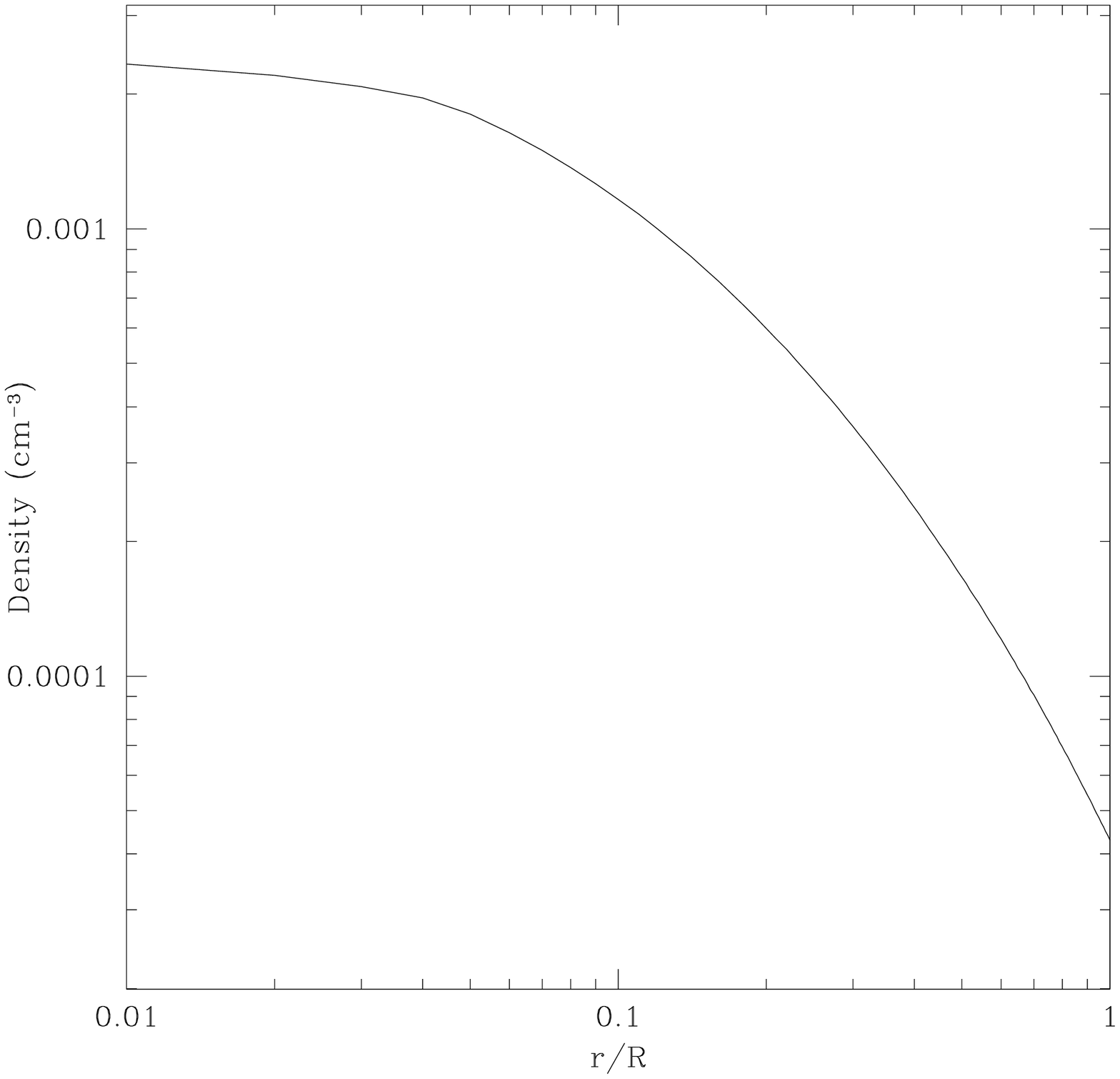}} \caption{\textit{left panel}:
Temperature profile that fits the projected emission-weighted
temperature profile of Fig.1; \textit{right panel}: Density
profile that fits the brightness profile of Fig. 1. \label{fig2}}
\end{figure*}



\clearpage



\begin{figure}
\centering \rotatebox{-00}{\epsscale{0.7} \plotone{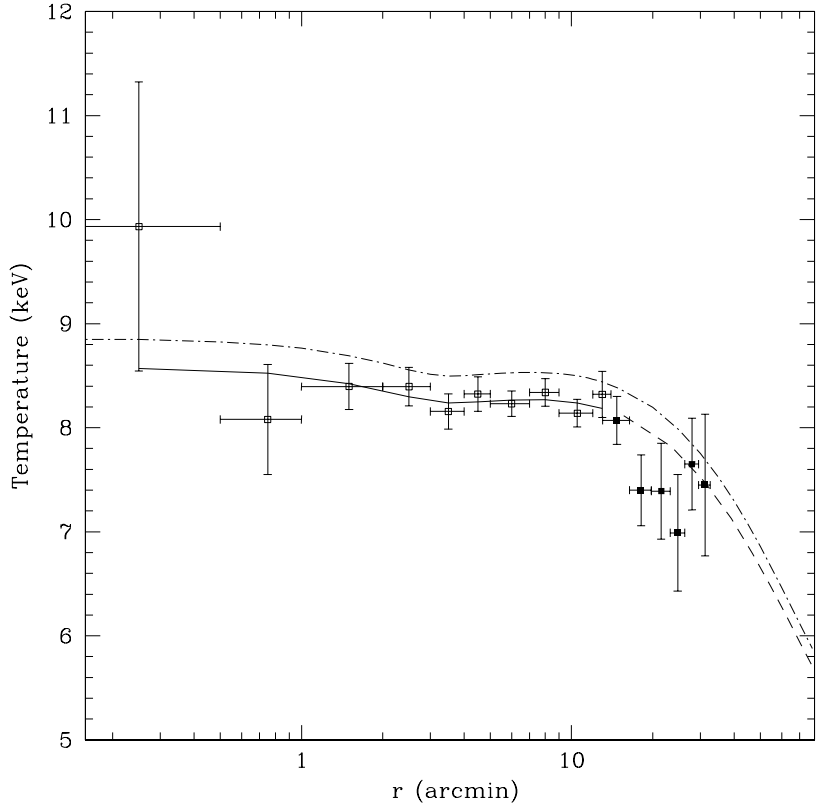}}
\caption{The continuous line represents the SM fit to the \xmm
temperature profile (empty square points, Snowden \etal 2008) and
the dashed line is the SM extrapolation to the virial radius (see
Fig. 1); the dot-dashed line is the SM fit with an average
temperature of 8.45 keV in the HXD-PIN FOV reported by the joint
\xmm/\suzaku analysis (W09). The filled square points are the
temperature values in the \xmm regions R1-R6 investigated by
W09.\label{fig3}}
\end{figure}

\clearpage

\begin{figure}
\centering \rotatebox{-90}{ \epsscale{0.7} \plotone{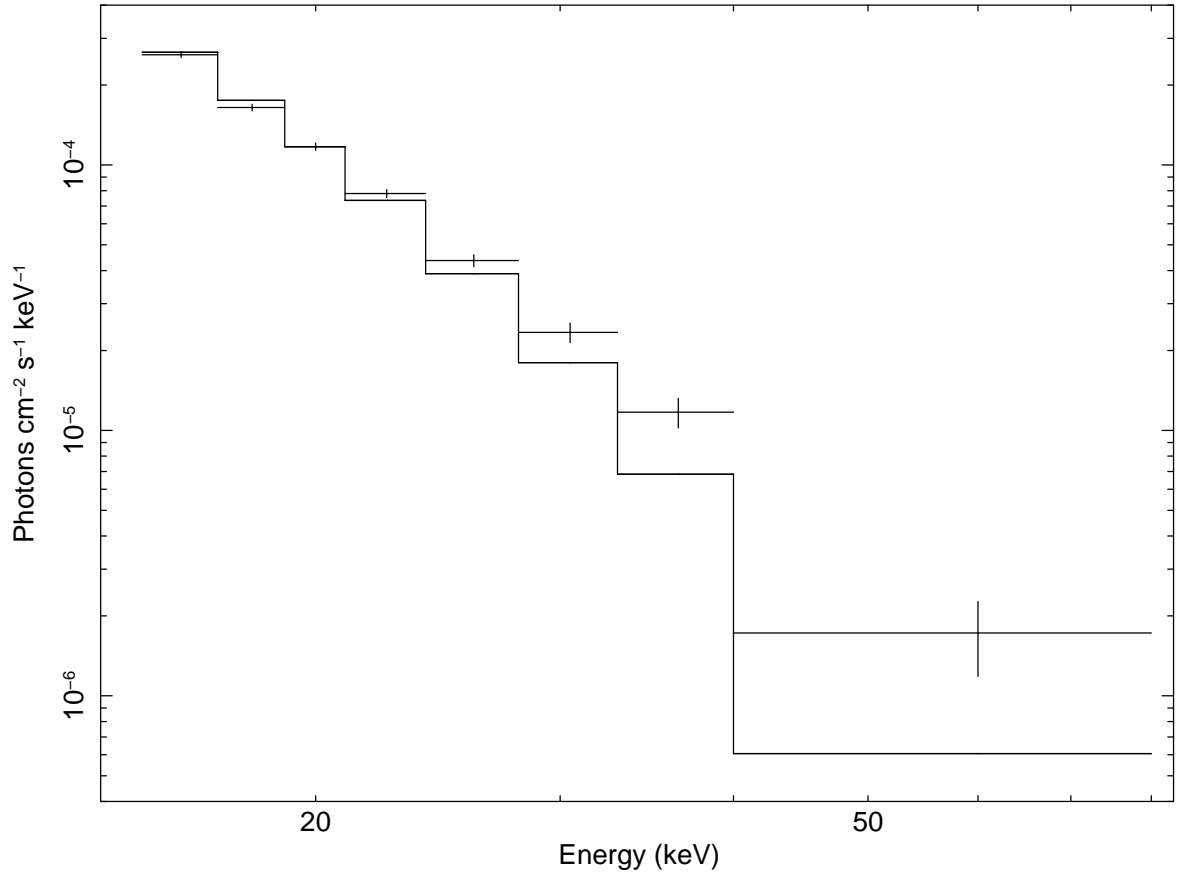}}
\caption{The data represent the HXR spectrum of the Coma cluster
observed by \bsax/PDS (FF04). The continuous line is the thermal
ICP emission derived from the SM analysis using the temperature
and density profiles of Fig. 2.\label{fig4}}
\end{figure}

\begin{figure}
\centering \rotatebox{-90}{\epsscale{0.7} \plotone{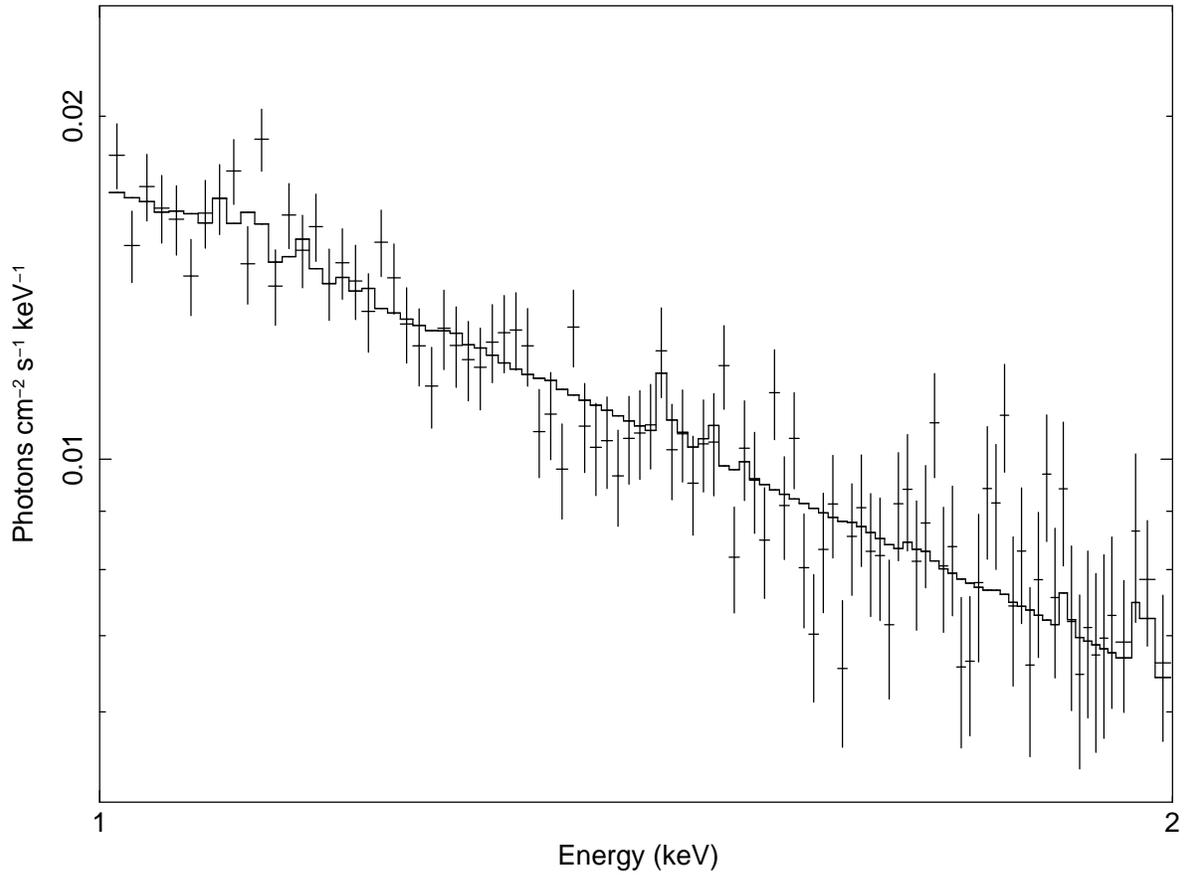}}
\caption{\rosat PSPC spectrum of the $20^{\prime}-40^{\prime}$
region around the center of the Coma cluster, fitted with the SM
profiles of Fig.2 ($\chi^2$=105.85/99). \label{fig5}}
\end{figure}

\end{document}